\documentclass{cseg2015}
\usepackage{xy}
\usepackage{amsmath,amssymb,graphicx,natbib}
\usepackage{rotating}
\usepackage{lscape}
\definecolor{burntorange}{RGB}{225,100,0}
\usepackage{setspace}
\usepackage{caption}
\usepackage{subcaption}

\begin{document}
 \vspace*{-2.cm}
\begin{center}
\includegraphics[height=3.2cm]{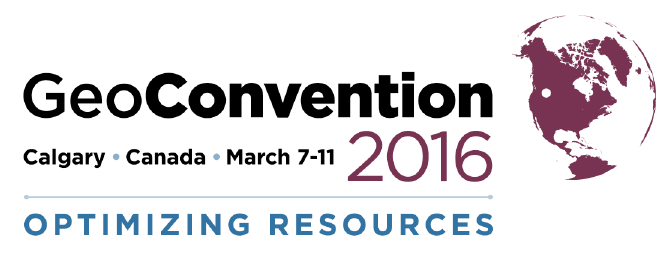}
\end{center}

{\LARGE {\bf Application of shifted-Laplace preconditioners for heterogenous Helmholtz equation{--} part 2: Full waveform inversion}} 


 \vspace*{-0.1cm}

{\it Nasser Kazemi}

\vspace*{-0.2cm}

{\it Department of Physics, University of Alberta}

\section{Summary}

Seismic waves bring information from the physical properties of the earth to the surface. Full waveform inversion (FWI) is a local optimization technique which tries to invert the recorded wave fields to the physical properties. An efficient forward-modelling engine along with local differential algorithm, to compute the gradient and Hessian operators, are two key ingredients of FWI approach. FWI can be done in time or frequency domain. Each method has its own pros and cons. Here, we only discuss frequency domain method with Krylov subspace solvers for time-harmonic wave equation. Nonlinearity of the problem requires good initial macro model of the physical properties and low frequency data. Macro models are built based on the kinematic information of the recorded wave fields. Another difficulty is the data modelling algorithm which is hard to solve especially for high wavenumbers (high frequencies). Without incorporation of high frequencies in the FWI algorithm we are not going to be able to update the macro models to high resolution ones. In the companion paper we showed that efficient forward modelling algorithms can be reached via proper preconditioners. Here, we will use the preconditioned data modelling engine in the context of local optimization method to solve for model parameters. The results show that we yield better convergence of the method and better quality of the inverted models after using the preconditioned FWI. 

\section{Introduction}

Improved computational resources, started a huge reinterest in the full wave form inversion (FWI) tools for recovering the physical properties of the earth \cite[]{Tarantola,Pratt1998,Virieux}. FWI is a local minimization method and can be implemented in time or frequency domain. Each method has its own pros and cons. Here, we use frequency domain algorithm. In frequency domain algorithms only limited number of frequency realizations needs to be inverted. Moreover, there is a possibility of inverting the frequency realizations with different grouping strategies to have a better convergence rate \cite[]{Amsalu}. One can also use the grouping strategies in linearized inversion problems of seismic imaging \cite[]{Kazemi_BLS}. However, the computational cost still is high and there is a problem of scalability of the algorithm that prevents usage of the method for bigger data sets and higher dimensions. FWI is a non-linear and ill-posed problem. Hence, we need to implement proper algorithms with some regularization terms. In this paper, we used Gauss-Newton method. Gauss-Newton method requires only the approximated version of the Hessian. Application of the approximated version of the Hessian in the context of least squares migration, for fast implementation, is also a common practice \cite[]{Hu,Kazemi_filter_based}. Another important factor in reducing the computational cost and scalability of the frequency domain FWI, is implementation of iterative algorithms for data modelling. Moreover, to increase the convergence rate one needs to apply proper preconditioners. In this paper we used shifted-Laplace preconditioners to increase the convergence rate of the data modelling engine in the context of local minimization method. Numerical examples on 1D and 2D synthetic data show the efficiency of the method in improving the convergence rate of the algorithm and the quality of the recovered physical properties of the subsurface.

\section{Full waveform Inversion in the space-frequency domain}
The Helmholtz equation with first-order absorbing boundary condition can be written as
\begin{equation}\label{eq:1}
 -\nabla . \rho \nabla {\bf u}-{k}^2(1-\hat{j} \alpha) {\bf u}={\bf f} \quad \mbox{in}\quad \Omega \in {\mathbb R}^d\quad d=1,2,3
 \end{equation}
where $\rho$ is density, ${k({\bf x},{\bf z})=\omega/c({\bf x},{\bf z})}$ is wavenumber and varies within $\Omega$ due to spatial variation of the velocity field $c({\bf x},{\bf z})$, $\omega$ is angular frequency related to the source function ${\bf f}$, $\alpha$ is damping parameter, ${\bf u}$ is pressure wave-field and ${\hat{j}=\sqrt -1}$ is complex identity. For constant density, equation (\ref{eq:1}) changes to
\begin{equation}\label{eq:2}
 -\Delta {\bf u}-\omega^{2}{\bf m}^{2}(1-\hat{j} \alpha) {\bf u}={\bf f}
 \end{equation}
where ${\bf m}=c({\bf x},{\bf z})^{-1}$ is real valued slowness model with dimension $M$. In matrix-vector notation equation (\ref{eq:2}) can be expressed as
\begin{equation}\label{eq:3}
{\bf B} {\bf u}={\bf f}
 \end{equation}
where ${\bf B}$ is impedance or Helmholtz matrix and ${\bf u}={\bf u} ({\bf m})$ is modelled wave field. Full waveform inversion is a local optimalization problem which tries to match the recorded wave field at the surface ${\bf u}_{obs}$ with the modelled data ${\bf u}_{cal}({\bf m})$ by solving
\begin{equation}\label{eq:4}
\hat{\bf m}=\underset{{\bf m}}{\operatorname{argmin}} \quad {\cal J}({\bf m})
 \end{equation}
 where ${\cal J}({\bf m})=||{\bf u}_{obs}-{\bf u}_{cal}({\bf m})||_2^2+\alpha\; |{\bf m}|_2^2$, ${\bf u}_{obs}$ and ${\bf u}_{cal}$ are measured and calculated wave fields, respectively, $\alpha$ is a regularization parameter and implicit summation is performed over the shots, receivers and frequency realizations of the wave fields (for more information interested readers are referred to \cite[]{Menke,Tarantola,Virieux}). Equation (\ref{eq:4}) is a non-linear problem and the minimum of the misfit functional is sought in the vicinity of the starting model ${\bf m}_0$. In other words we have ${\bf m}={\bf m}_0+\Delta {\bf m}$ where $\Delta {\bf m}$ is a perturbation model. The starting model is built based on the kinematic information of the recorded wave field. Different algorithms can be used to solve for the updated model (e.g., gradient descent, Gauss-Newton, Full Newton). Here, we only discuss Gauss-Newton method. We will follow the matrix-vector notations of \cite{Pratt1998}.

To solve the problem let's expand the misfit functional via Taylor series up to second order
\begin{equation}\label{eq:5}
{\cal J}({\bf m}_0+\Delta {\bf m})={\cal J}({\bf m}_0)+\Delta {\bf m}^T\nabla_m {\cal J}({\bf m}_0)+\frac{1}{2}\Delta {\bf m}^T{\bf H}\Delta {\bf m}+{\cal O}(|\Delta {\bf m}|^3)
 \end{equation}
where $H$ is the Hessian and $T$ is transpose operator. Taking the first order derivative of the misfit functional with respect to the model parameters and seeking the first order optimality condition we get
\begin{equation}\label{eq:6}
\Delta {\bf m}=-{\bf H}^{-1}\nabla_m {\cal J}({\bf m}_0)
\end{equation}
where

\begin{equation}\label{eq:7}
\nabla_m {\cal J}({\bf m}_0)={\cal R}[(\frac{\partial {\bf u}_{cal}({\bf m}_0)}{\partial {\bf m}})^T({\bf u}_{obs}-{\bf u}_{cal}({\bf m}_0))]+\alpha {\bf m}_0={\cal R}[{\bf G}_0^T\Delta {\bf u}_0]+\alpha {\bf m}_0
\end{equation}
and differentiating the gradient term of equation (\ref{eq:7}) with respect to the model parameters results in

\begin{equation}\label{eq:8}
{\bf H}=\frac{\partial^2 {\cal J}{\bf m}_0}{\partial^2 {\bf m}}={\cal R}[{\bf G}_0^T{\bf G}_0+\alpha {\bf I}]+{\cal R}[(\frac{\partial {\bf G}_0}{\partial {\bf m}})^T(\Delta {\bf u}_0^*\dots \Delta {\bf u}_0^*)]
\end{equation}
where ${\cal R}$ is for real part, $*$ denotes conjugate of complex number and ${\bf G}$ is the sensitivity or the $\mbox{Frech\'et}$ derivative matrix and ${\bf I}$ is an identity matrix.
In Gauss-Newton method we drop the second term from the Hessian matrix and use only the approximate version of it. This term counts for changes in the gradient vector due to the second order non-linear effects (e.g., first order multiples). By dropping the second term from the Hessian matrix and plugging in equations ({\ref{eq:7}}) and ({\ref{eq:8}}) into the equation ({\ref{eq:6}}) we get
\begin{equation}\label{eq:9}
{\bf m}={\bf m}_0-{\cal R}[{\bf G}_0^T{\bf G}_0+\alpha {\bf I}]^{-1}\;({\cal R}[{\bf G}_0^T\Delta {\bf u}_0]+\alpha {\bf m}_0).
\end{equation}
 In each iteration of the Gauss-Newton method, we need to calculate three forward modellings; one for calculating the data residual based on the updated model from previous iteration and two forward modellings for back propagating  the data residual to calculate ${\bf G}_0$ and ${\bf G}_0^T$. In this paper, iterative data modelling via Krylov subspace method in frequency domain is implemented. We also applied shifted-Laplace operators to precondition the system of equations and then GMRES method is used to model the data (for more information see \cite[]{Gijzen,kazemi-data-modelling}). In comparison to direct solvers, Krylov subspace algorithm has better scalability potential for bigger geometries and data sets. Moreover, as it is clear from equation (\ref{eq:9}), we need to calculate the inverse of the Hessian matrix and in practice Hessian matrix is dense, therefore Conjugate Gradient method is used to update the model perturbation. In next section we will show the degree of usefulness of the preconditioners in improving the convergence rate of Gauss-Newton method and the quality of the recovered physical properties of the earth in the context of the space-frequency domain full waveform inversion. 
  \vspace{-0.1cm}
\section{Examples}

\begin{figure}
\centering
\begin{minipage}{.5\textwidth}
  \centering
  \vspace{-1.45 cm}
  \includegraphics[width=.7\linewidth]{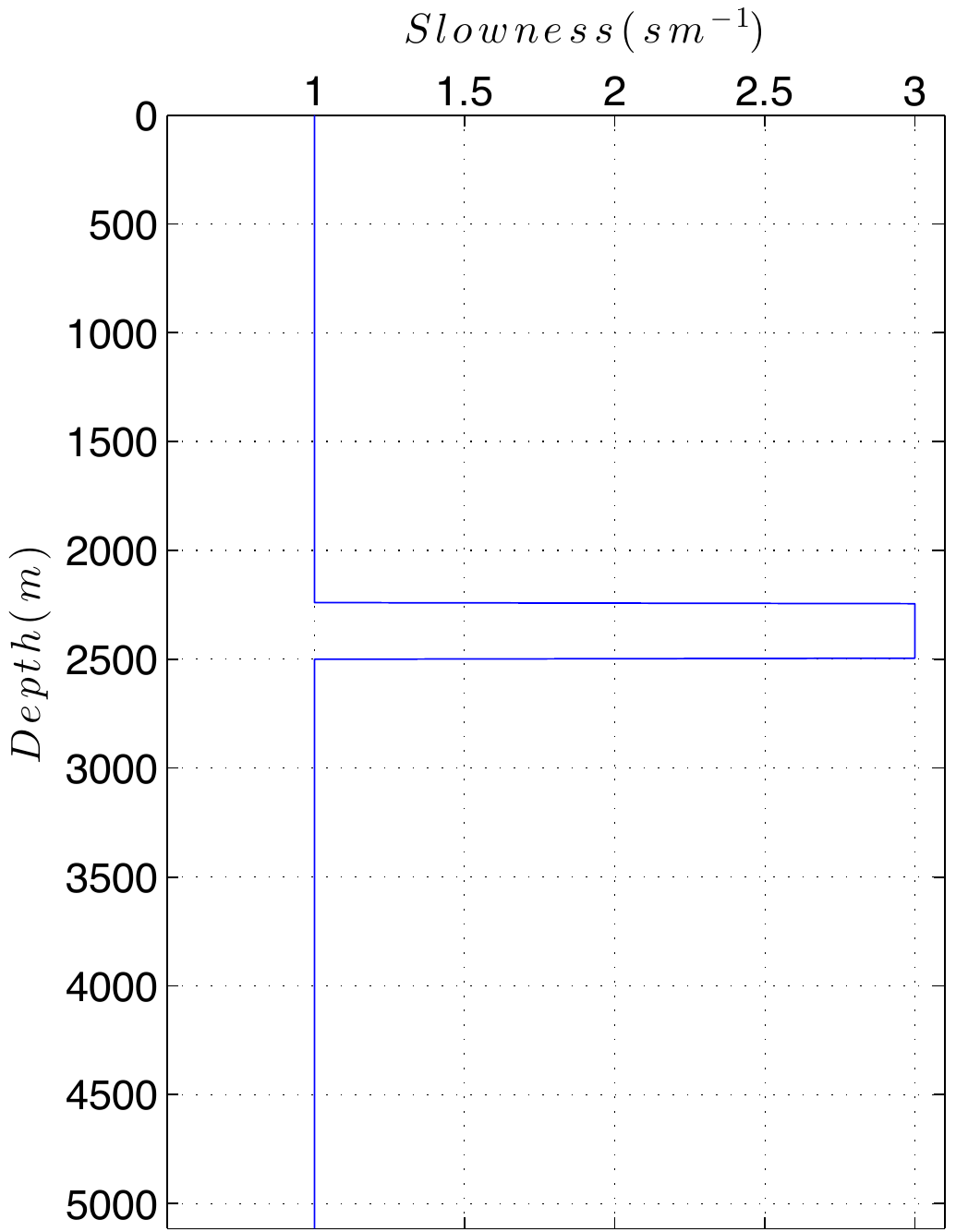}
  \captionof{figure}{True 1D scattering wave field model.}
  \label{fig:1}
\end{minipage}%
\begin{minipage}{.5\textwidth}
  \centering
  \includegraphics[width=.7\linewidth]{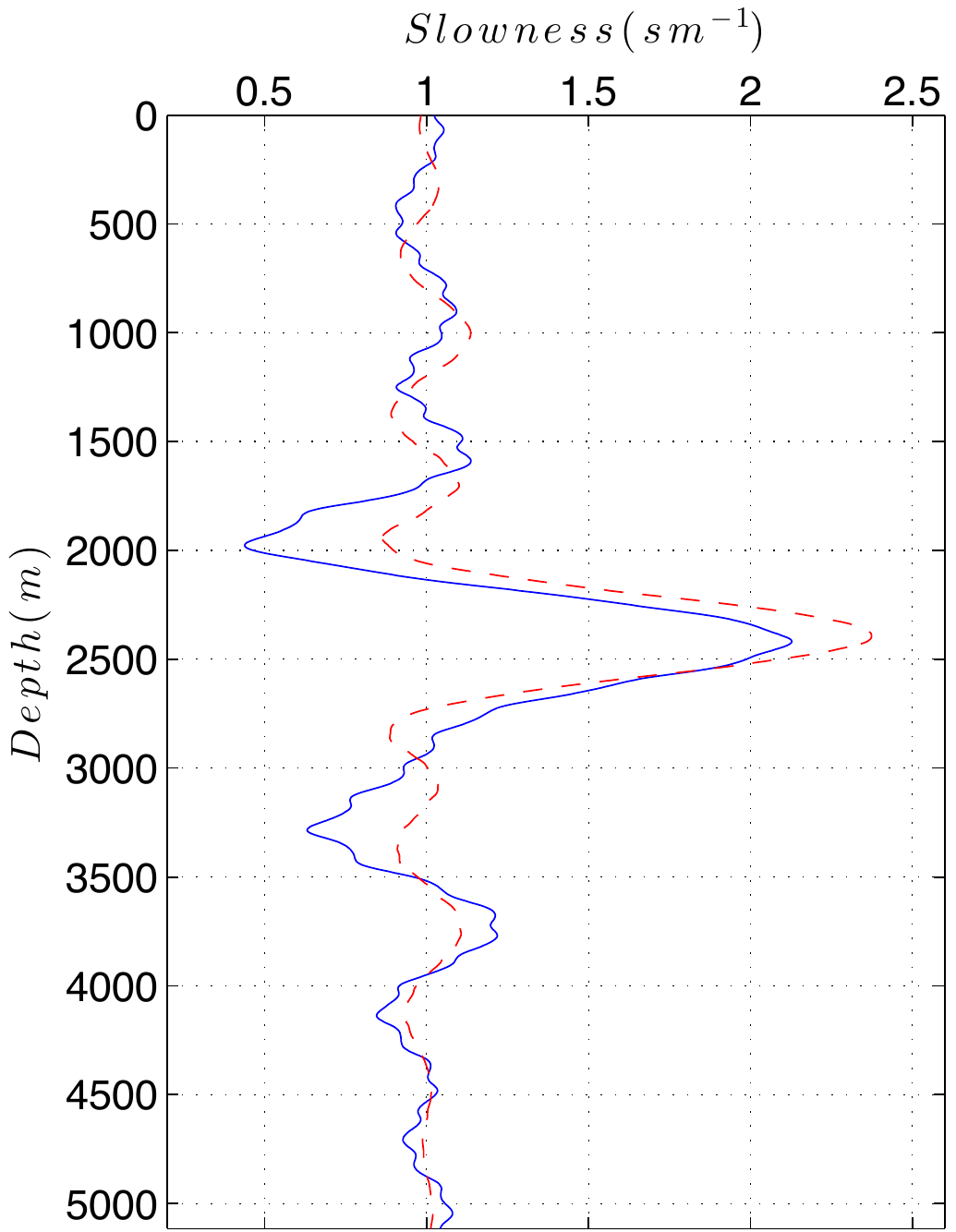}
  \captionof{figure}{Inverted models using Gauss-Newton method for the 1D example in Figure \ref{fig:1} (Red colour with preconditioner and blue colour without preconditioner).}
  \label{fig:2}
\end{minipage}
\end{figure}
\begin{figure}[]
  \vspace{-0.1cm}
  \begin{center}
    \includegraphics[width=0.5\textwidth,height=8cm]{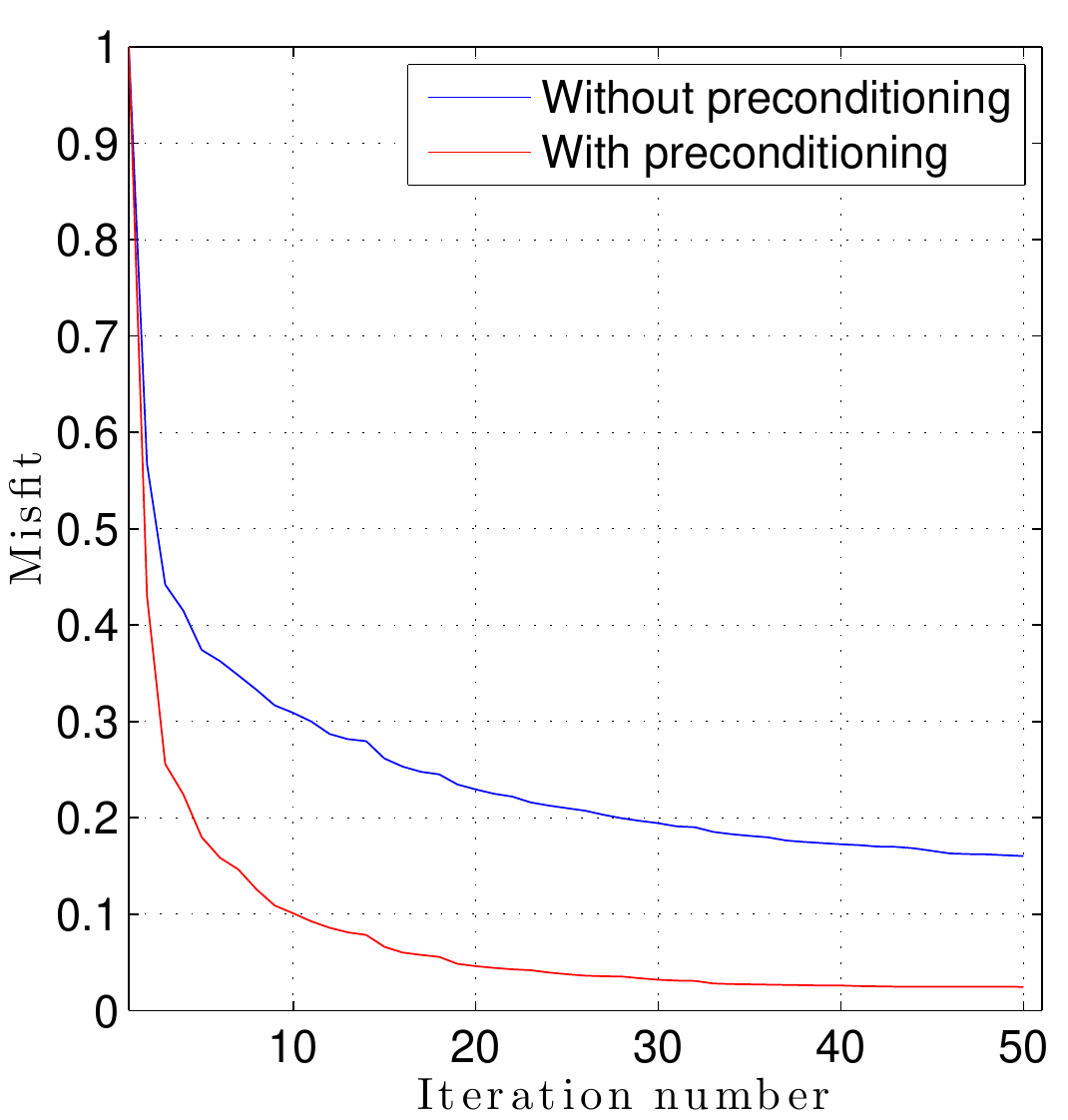}
  \end{center}
  \caption{Convergance behaviour of the Gauss-Newton method for the inverted models in Figure \ref{fig:2}. }
  \label{fig:3}
  \end{figure}

\vspace{-.1 cm}
To demonstrate the method we generated a 1D model (Figure \ref{fig:1}). Sources and receivers are placed along the Well. The source interval is $10m$ and receiver interval is $5m$. We generated the measured data set by doing frequency domain finite difference algorithm via LU factorization (0.2, 0.5, 1, 5 and 10$Hz$ frequencies are used). To solve for the physical properties of the subsurface, we run Gauss-Newton method with and without preconditioners. All of the frequencies are inverted at the same time (For different frequency selection strategies see \cite{Amsalu}). We used fixed mesh size for all of the frequency realizations. For the preconditioner, we used shifted-Lplace operators with complex shift equal to  $1+\hat{j}$. As we mentioned before, for the data modelling part of the algorithm, we implemented Krylov subspace algorithm. GMRES method with and without preconditioners are used. Figure \ref{fig:2} shows the inverted models using the Gauss-Newton method with and without preconditioners. The inverted model is improved after applying the proper preconditioner to the data modelling engine. Because, we are using the same engine to calculate the gradient and Hessian operators, one should expect the improvement in the convergence of the algorithm as well. The convergence behaviour of the method before and after applying the shifted-Laplace preconditioners are shown in Figure \ref{fig:3}. As it is clear from the figure, the convergence rate of the method drastically increased after applying the preconditioner.


\begin{figure}[]
  \vspace{-0.1cm}
  \begin{center}
    \includegraphics[width=1\textwidth]{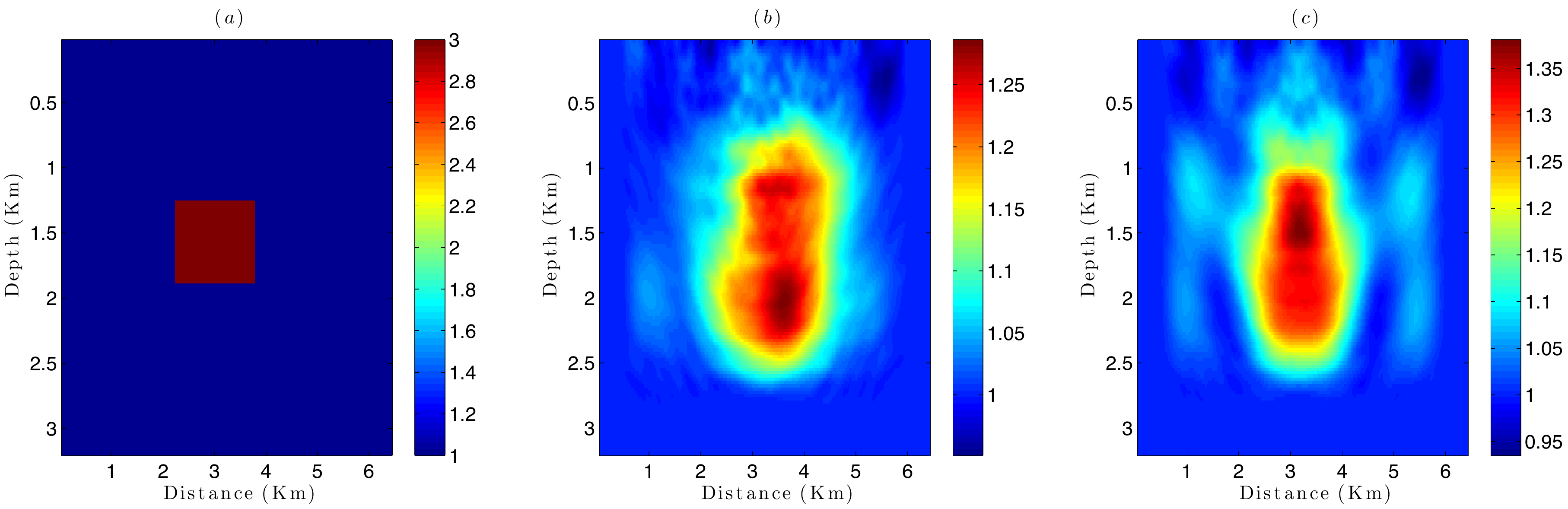}
  \end{center}
  \caption{True and inverted models for the 2D example (0.5, 2 and 5$Hz$ frequencies are used). a) True slowness model. b) Inverted model using Gauss-Newton without preconditioner. c) Inverted model using Gauss-Newton with preconditioner.}
  \label{fig:4}
  \end{figure}
  
  \begin{figure}[]
  \vspace{-1 cm}
  \begin{center}
    \includegraphics[width=0.5\textwidth,height=8cm]{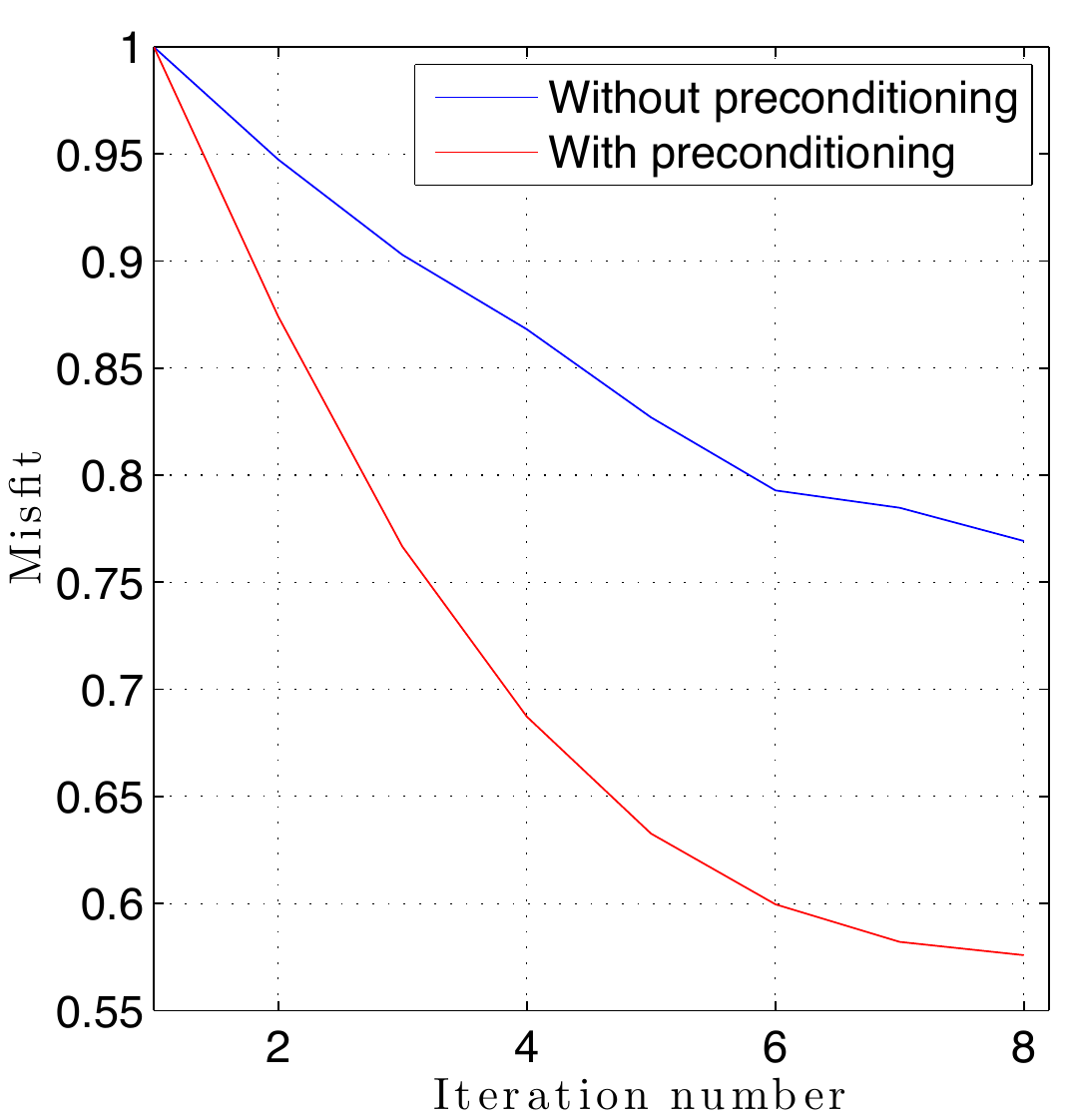}
  \end{center}
  \caption{Convergance behaviour of the Gauss-Newton method for the inverted models in Figure \ref{fig:4}b and c. }
  \label{fig:5}
  \end{figure}
Next, we applied the method on 2D model (Figure \ref{fig:4}a). The source and receivers are deployed at the surface with receiver intervals of $25m$ and only $10$ shots are fired on a whole survey with equal spacing. Again we generated the measured data set by doing frequency domain finite difference algorithm via LU factorization (0.5, 2 and 5$Hz$ frequencies are used). Gauss-Newton method is used to solve for the subsurface model. Here, we invert the frequencies in a sequential order. We first invert for the lowest frequency and then use the updated model as an initial estimate for the next frequency realization. We did not change the mesh size for the different frequency realizations. The results of inverted models with and without preconditioners are shown in Figures \ref{fig:4}b and c,  respectively. Because of the limited number of shots and frequency realizations there is smearing in the final solutions. However, after the implementation of the shifted-Laplace preconditioner we were be able to improve the performance of the algorithm. Convergence behaviour of the Gauss-Newton method with and without preconditioner is shown in Figure \ref{fig:5}. As it is clear from the figure, the convergence rate of the method greatly increased after applying the preconditioner.

\section{Conclusion}
Preconditioned FWI in space-frequency domain is implemented and the results are compared with the case that we did not use preconditioning. For data modelling part of the Gauss-Newton technique, Krylov subspace algorithm (GMRES) is incorporated. Shifted-Laplace operator proved to be a valid candidate for preconditioning the frequency domain FWI problem. Numerical examples on 1D and 2D synthetic data showed the efficiency of the method in improving the convergence rate of the algorithm and the quality of the recovered physical properties of the subsurface.

 \vspace{-0.1cm}
\section{Acknowledgements}
The authors are grateful to the sponsors of Signal Analysis and Imaging Group (SAIG) at the University of Alberta.
\bibliography{paper-fwi.bib}

\end{document}